# An Evolvable Fuzzy Logic System for handoff management in heterogeneous Wireless Networks

Hossein Fayyazi
Faculty of ICT, Malek-Ashtar University of Technology
Tehran, Iran
E-mail: fayyazi_hossein@yahoo.com

Mohammad Sabokrou
Faculty of ICT, Malek-Ashtar University of Technology
Tehran, Iran
E-mail: sabokro@gmail.com

*Abstract*—One of the features of the Next Generation Wireless Networks (NGWNs) is its heterogeneous communication environment. Heterogeneous networks are ranging from wireless WAN, LAN, MAN and PAN. The most important parameters in this regard are different data rates and coverage of these networks. In this paper, we propose an evolvable fuzzy logic-based algorithm for managing handover process in heterogeneous wireless networks. We use fuzzy logic system to determine the right time of initiating the handoff procedure and utilize Genetic Algorithm (GA) to predict the best form of rules of the fuzzy module. We implement our proposed scheme on a special simulation model and discuss about the impact of evolution on the algorithm.

*Keywords-heterogeneous wireless networks; fuzzy logic system; Genetic Algorithm; handoff*

## I. INTRODUCTION

Next Generation Wireless Systems (NGWS) will provide a variety of services to mobile users, including high-speed data, real-time applications and multimedia support [1]. Since there is no one single wireless network technology that can satisfy the requirements of all today's and upcoming wireless services, the coexistence of heterogeneous wireless networks to provide service anywhere at any time is an inevitable trend in the development of the NGWNs [2].

One of the most important parts of any mobile communication network is the handoff management procedure. Transition of an active connection from one Base Station (BS) to another one is called handover process. In the handover process, the new BS assigns one of its unoccupied channels to the Mobile Terminal (MT) while the connection of MT with the old BS remains active for some time slots.

Appointment of the exact time of initiating the handover process is an important issue. Handoff decision is based on the Received Signal Strength (RSS) [2]. The handoff procedure should be carried out successfully before the MT moves out of the coverage area of the old BS.

In this paper, we propose a fuzzy logic-based handoff management system and use GA as the predictor of the consequences of the fuzzy rules so that the minimum number of handoffs occurs and the total number of cut connections decrease.

We use the MT's velocity and the distance of the MT with the boundary of the BS and the number of free channels of it as inputs of our fuzzy system and the RSS threshold will be the output. When this threshold drops below a predetermined value $S_{th}$ then the handover process is initiated and when it drops below $S_{min}$, it shows that the MT moves beyond the coverage area of the BS and disconnected from it.

We suppose that the information used as inputs of the fuzzy system are available from different sources. For example, speed information is derived from link layer of the TCP/IP protocol stack that uses VEPSD [3] to estimate the mobile's speed.

The rest of the paper is organized as follows. A review of related works of handoff decision in NGWS is given in section II. Section III provides the necessary background information about handover and different types of it and section IV gives a description of the proposed scheme for handoff management. We test the proposed method on a special simulation model that section V introduces it. The experimental results and conclusion are provided in VI and VII, respectively.

## II. RELATED WORKS

A survey on mobility management in NGWS is presented in [4]. Reference [5] enhances the handoff performance of mobile IP in wireless IP networks by reducing the false handoff probability in the NGWS handoff management protocol and analyzes the false handoff probability's effect on mobile speed and handoff signaling delay. A fuzzy logic-based handoff management protocol for NGWS is proposed in [6], which is integrated with an existing cross-layer handoff protocol. Reference [7] offers a handoff management architecture using relative signal strength of the old and new BSs to calculate the handoff initiation time. A GPS-based handoff technique for improving the handoff probability in NGWS is presented in [8]. The GPS helps determine the direction of the MT and so reduces the false handoff probability. Reference [9] tries to minimize the handoff latency with a new scanning method in which determines the distance of the nearest BS from the MT to bypass the main processes involved in increasing MAC layer handoff latency. In [1] a genetic-based call admission control scheme is suggested.

## III. INTRODUCTION TO HANDOVER

NGWS integrate various existing wireless network technologies; each is optimized for some specific services such as WLANs, WiMax, General Packet Radio Service and Universal Mobile Telecommunication System [7]. These different networks overlap each other hierarchically and a multi-interface Mobile Terminal can select an appropriate network to use [2].

The cellular concept is the idea of replacing large, single high-power transmitter cells with several small, low power ones. Each of these cells typically provides coverage to a small portion of the coverage area [10].A cell is the radio area covered by a transmitting station or a BS [5] which the base station transmitter is placed at the centre of the cell and services all the MTs within the cell area [10].

Handover is the process of transferring an active MT session from one BS or Access Point (AP) to another one, in order to keep the user's connection uninterrupted [11]. The connection between a mobile user and a BS or AP remains active until the RSS is upper than an admissible threshold ($S_{th}$). When RSS is decreased continuously and reaches below this threshold, the session must be switched to a neighboring cell. If the neighboring cell belonged to the same system of the current cell, handover between these cells is termed as Horizontal Handoff. Handover between two BSs that belong to two different network interfaces is called Vertical Handoff.

Handoff can be done in two ways, soft and hard. In hard handoff, the MT must be disconnected from the old BS and then connected to the new one. Unlike the hard handoff, the connection to the old BS is not broken until the MT is connected to the new BS. In other words, in soft handoff, the MT is connected to more than one BS simultaneously. Fig 1 shows these concepts.

Presenting an admissible management protocol for the integrated architecture of NGWS is an important and challenging issue. Handoff management protocols operate in different layers of the TCP/IP protocol stack (e.g., link layer, network layer, transport layer, and application layer) [12].

## IV. PROPOSED SCHEME

### A. Fuzzy Module

We use velocity, distance and number of free available channels as input and RSS threshold as output of the fuzzy-based part of our algorithm and exploit GA as the predictor of the consequences of the fuzzy rules.

Figs 2 – 5 shows the Membership Functions (MFs) used in proposed scheme and the initial rule set has been presented in Table I. The MFs of the number of free channels fuzzy variable vary in the execution time according to its respective BS, because the number of channels varies in different BSs.

This module utilizes the Mamdani fuzzy logic system and Center of Area defuzzification method.

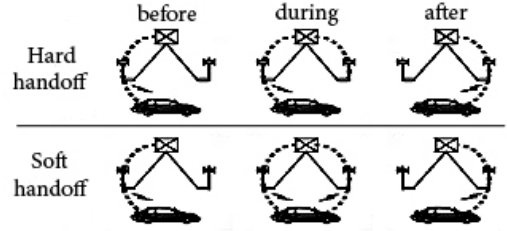

Figure 1. Hard handoff VS. Soft handoff

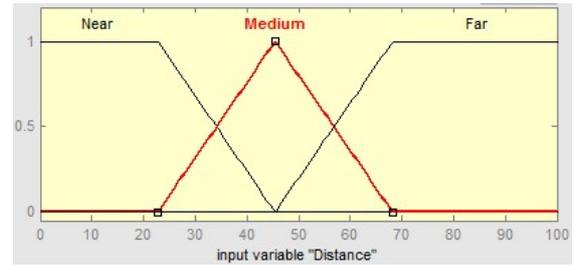

Figure 2. Membership function of fuzzy variable distance

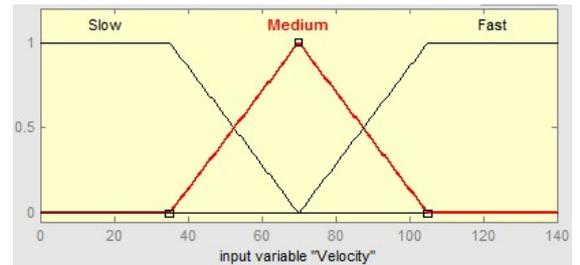

Figure 3. Membership function of fuzzy variable velocity

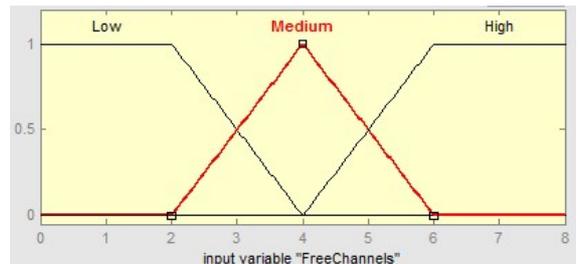

Figure 4. Membership function of fuzzy variable number of free channels

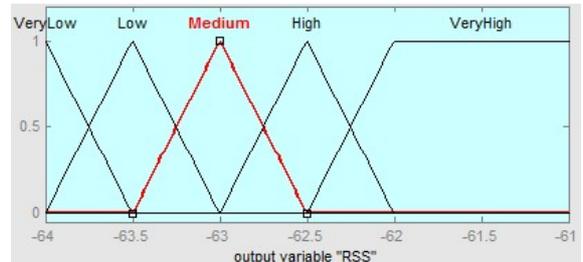

Figure 5. Membership function of fuzzy variable RSS

TABLE I. RULE SET OF PROPOSED FUZZY LOGIC SYSTEM

| Velocity | Distance | Number of free channels | RSS thr. |
|---|---|---|---|
| Slow | Near | Low | Low |
| Slow | Near | Medium | Low |
| Slow | Near | High | Medium |
| Slow | Medium | Low | Medium |
| Slow | Medium | Medium | Medium |
| Slow | Medium | High | High |
| Slow | Far | Low | High |
| Slow | Far | Medium | Very High |
| Slow | Far | High | Very High |
| Medium | Near | Low | Very Low |
| Medium | Near | Medium | Low |
| Medium | Near | High | Low |
| Medium | Medium | Low | Medium |
| Medium | Medium | Medium | Medium |
| Medium | Medium | High | Medium |
| Medium | Far | Low | High |
| Medium | Far | Medium | High |
| Medium | Far | High | High |
| Fast | Near | Low | Very Low |
| Fast | Near | Medium | Very Low |
| Fast | Near | High | Low |
| Fast | Medium | Low | Low |
| Fast | Medium | Medium | Low |
| Fast | Medium | High | Medium |
| Fast | Far | Low | Medium |
| Fast | Far | Medium | High |
| Fast | Far | High | High |

*B. Evolutionary Module*

To reduce handoffs occurring in the proposed method, we use GA as a subsidiary tool. GA updates the consequences of the fuzzy rule set and tries to change them so that the total number of cut connections and handoffs decrease.

The chromosome representation used in this approach is as {Consequence$_1$, …, Consequence$_n$} where n is the total number of rules (here, n = 27). As you see in Fig 5, we consider 5 possible values for RSS threshold, "Very Low", "Low", "Medium", "High", "Very High" and so there are 5 possible integer values from 1 to 5 for consequences in chromosome representation.

In our approach, the population size is set to 50 and one-point crossover and random-resetting mutation are used. The crossover and mutation probabilities are set to 0.9 and 0.1, respectively. The selection mechanism is tournament with size of 10.

To evaluate each chromosome, we run the fuzzy module for the four latest units of time of the simulation model and count the number of handoffs and cut connections. The chromosomes with the minimum number of handoffs and cut connections will be the best.

V. SIMULATION MODEL

The simulation environment is considered as follows. 7 BSs cover an environment with 6000×6000 m$^2$ area. At the first unit of time, 50 MTs with different velocities and directions start moving from different places. Velocity, direction and the start position of all MTs are determined, randomly. To apply the heterogeneous nature of wireless networks, the number of channels and the radius covering environment of each BS are different from the others. The coordinates, radiuses and number of channels vectors assumed for the BSs are {(2598, 500), (866, 500), (3464, 2000), (1732, 2000), (1, 2000), (2598, 3500), (866, 3500)}, {1400, 1000, 1200, 800, 900, 600, 1300} and {6, 4, 5, 3, 3, 2, 5}, respectively. We analyze the connection time, number of handoffs and energy wastage of these MTs in 75 units of time. Fig 6 represents one MT which is moving through the simulation environment and is connected to two BSs because of the handover process.

If the movement type of the MT is to be accelerated, then the velocity ($v_i$) and the place of terminal ($x_i$) in each unit of time ($t_i$) is calculated using equations (2),(3) in which Δx is the total distance and t is the total time units.

$$a = \frac{2\Delta x}{t^2} \quad (1)$$
$$v_i = \sqrt{2\,a\,t_i} \quad (2)$$
$$x_i = \frac{1}{2} a t_i^2 \quad (3)$$

On the other hand, if the MT moves with a steady velocity, then the place of the MT in each unit of time can be obtained from equation (4).

$$x_i = v_i t_i \quad (4)$$

We consider 100 energy units for each MT at the beginning of the simulation. This energy decreases as MT is connected to different BSs at simulation time. The more distance from the BS leads to the more reduction in energy and so the energy wastage in MT is based on its distance from the serving BS. Equation (5) shows this relevance.

$$EW = \frac{d_{BS}^{MT}}{r_{BS}} + \varepsilon \quad (5)$$

Where $d_{BS}^{MT}$ is the distance of MT from its serving BS, $r_{BS}$ is BS's radius and ε is a constant value (e.g. ε = 0.1).

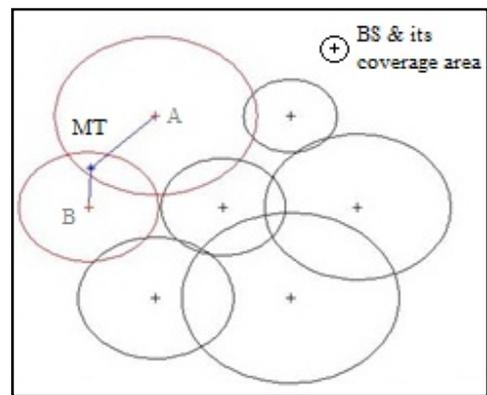

Figure 6. The simulation environment with one MT in handover status. The MT is connect to A and B BSs, simultaneously.

Each MT has one of the "connect", "disconnect" or "Handover" states in each unit of time and four actions can be applied to them. Table II shows these concepts. The "Next state" column in Table II occurs when the respective action is accomplished successfully. As shown in this table, in the handoff procedure the connection between the old BS, and the new one remains active in some units of time, where in one time-unit the "initial handover" action converts the connect state to handover and tries to connect to the new BS and in the other, the "end handover" action converts this handover to connect state and the connection of the old BS is cut. In our simulation model, the connection remains active in the two BSs for two-time units. This delay helps the system to avoid the "ping pong" effect.

TABLE II. DIFFERENT STATES AND ACTIONS APPLIED TO MTS

| RSS thr VS $S_{th}$ | RSS thr VS $S_{min}$ | Current state | Action | Next state |
|---|---|---|---|---|
| RSS thr $<$ $S_{th}$ | RSS thr $>$ $S_{min}$ | Connect | Initial handover | Handover |
| - | RSS thr $<$ $S_{min}$ | Connect | Cut connection | Disconnect |
| - | RSS thr $>$ $S_{min}$ | Disconnect | Try to connect | Connect |
| - | RSS thr $>$ $S_{min}$ | Handover | End handover | Connect |

## VI. SIMULATION RESULTS

We implement the simulation model introduced in previous section and apply the fuzzy module of handoff management protocol proposed in [6] as FLAH and our proposed method with GA as GFLS to compare their results. We also apply the Evolutionary module to FLAH as GFLAH. Table III represents the results of simulation model that is executed for 10 times.

TABLE III. SIMULATION RESULTS

| Algorithm | | FLAH | GFLAH | GFLS |
|---|---|---|---|---|
| Number Of Handoffs | Max | 118 | 53 | **33** |
| | Min | 30 | 27 | **14** |
| | Avg | 58 | 38 | **23** |
| Connection Time (Percentage) | Max | 28.70% | 29.41% | **31.62%** |
| | Min | 24.49% | 24.70% | **26.82%** |
| | Avg | 27.00% | 27.35% | **29.38%** |
| Energy wastage (Percentage) | Max | 19.56% | **17.75%** | 18.59% |
| | Min | **13.83%** | 14.36% | 15.33% |
| | Avg | 16.45% | **16.17%** | 16.76% |

## VII. CONCLUSION AND FUTURE WORKS

From the simulation results, evolution can manage the variable conditions of the environment truly and gives good results. FLAH does not consider the number of free channels of each BS as an input fuzzy variable and so the combination of FLAH with evolution module enhances the quality of the management scheme considerably. Therefore, in real world scenarios with different variable parameters like handoff latency, minimum network cost, angle of MT's motion, etc, exploiting evolution has a great impact on the quality of handoff management procedure.

In future works, other parameters that are more common in real scenarios may be evaluated and included in inputs of fuzzy module to improve system performance.